\documentclass[11pt]{article}

\usepackage{epsfig}
\usepackage{amsfonts,amssymb}

\newcommand{\al}{\ensuremath{\alpha}}

\newcommand{\ep}{\ensuremath{\epsilon}}

\newcommand{\ka}{\ensuremath{\kappa}}
\newcommand{\la}{\ensuremath{\lambda}}

\newcommand{\Om}{\ensuremath{\Omega}}
\newcommand{\p}{\ensuremath{\phi}}

\renewcommand{\t}{\ensuremath{\tau}}
\renewcommand{\th}{\ensuremath{\theta}}

\font\mybb=msbm10 at 12pt 
\def\bb#1{\hbox{\mybb#1}}

\def\bE {\bb{E}} 
 
\def\bM {\bb{M}}

\newcommand{\D}{\ensuremath{{\cal D}}}
\newcommand{\E}{\ensuremath{{\cal E}}}

\newcommand{\T}{\ensuremath{{\cal T}}}

\newcommand{\ra}{\ensuremath{\rightarrow}}
\newcommand{\del}{\ensuremath{\partial}}

\newcommand{\td} {\ensuremath{\tilde}}
\newcommand{\inv}{\ensuremath{^{-1}}}
\newcommand{\half}{\ensuremath{\frac{1}{2}}}

\newcommand{\be}{\begin{equation}}
\newcommand{\ee}{\end{equation}}

\newcommand{\ba}{\begin{eqnarray}}
\newcommand{\ea}{\end{eqnarray}}

\title{{\bf A note on the Supergravity Description of Dielectric Branes}}
\author{{\bf D. Brecher}\thanks{email: dominic.brecher@durham.ac.uk}~~and
{\bf P. M. Saffin}\thanks{email: p.m.saffin@durham.ac.uk}
\\ Centre for Particle Theory, Department of Mathematical Sciences, \\
University of Durham, South Road, Durham DH1 3LE, 
United Kingdom.\\ \\ Preprint DCPT-01/51}

\date{2 July 2001}
	
\begin{document}

\maketitle

\begin{abstract}

We comment on the recent papers by Costa \emph{et al} and Emparan,
which show how one might generate supergravity solutions describing
certain dielectric branes in ten dimensions.  The ``basic'' such
solutions describe either $N$ fundamental strings or $N$ D4-branes
expanding into a D6-brane, with topology $\bM^2 \otimes S^5$ or $\bM^5
\otimes S^2$ respectively.  Treating these solutions in a unified way,
we note that they allow for precisely two values of the radius of the
relevant sphere, and that the solution with the smaller value of the
radius has the lower energy.  Moreover, the possible radii in both
cases agree up to numerical factors with the corresponding solutions
of the D6-brane worldvolume theory.  We thus argue that these
supergravity solutions are the correct gravitational description of
the dielectric branes of Emparan and Myers.

\end{abstract}

\newpage

\section{Introduction}

The higher-dimensional counterparts of the Melvin
universe~\cite{melvin:64}, which describe gravitating
magnetic fluxbranes, have appeared in various string- and M-theoretic
guises over the last few years 
(see, e.g., \cite{gibbons:88}--\cite{paul2}).  
Within the Kaluza-Klein context, the
most natural way to generate such fluxbranes is via a twisted
compactification of flat space~\cite{dowker:93,dowker:95a,dowker:95}.  Moreover, starting
instead with the Euclidean Schwarzschild solution cross a
trivial time direction, one can generate \emph{spherical} branes via
just such a twisted compactification~\cite{dowker:95}.  In slightly different
language, one can think of this as follows: take a $D$-dimensional
uncharged black string; analytically continue the worldvolume
coordinates; and perform a double dimensional reduction along a twisted
direction.  The resulting $(D-1)$-dimensional solution describes a
spherical $(D-5)$-brane in the core of a magnetic
fluxbrane.  It has local, but zero net, magnetic charge\footnote{More
generally, spherical $(D-3-2n)$-branes in the core of $n$ intersecting
fluxbranes can be generated by applying $n$ twists~\cite{dowker:95}.}.

If one now adds $N$ units of 2-form charge to the $D$-dimensional black string, the
$(D-1)$-dimensional solution will exhibit $N$ units of 1-form charge, so is
more rightly interpreted as a collection of $N$ particles ``expanding'' into a sphere
under the influence of the background $(D-3)$-form magnetic field.  Of
course, these techniques need not only be applied to black
\emph{strings} and, in recent work, Costa \emph{et al}~\cite{costa:01}
and Emparan~\cite{emparan:01} considered their application in an
M-theoretic context.  Starting with the basic branes of
eleven-dimensional supergravity, then, these authors showed how to
generate solutions of type IIA supergravity which
describe various ten-dimensional branes expanding into a D6-brane
under the influence of a background 8-form magnetic field.  In
particular, the reduction of certain M2- or M5-brane solutions generates
ten-dimensional solutions describing F-strings or D4-branes expanding into a D6-brane
with topology $\bM^2 \otimes S^5$ or $\bM^5 \otimes S^2$
respectively, where $\bM^n$ denotes an $n$-dimensional
Minkowski space.  Applying T-duality generates more general configurations~\cite{emparan:01} --
D$p$-branes expanding into a D$(p+2)$-brane, and F-strings expanding
into a D$p$-brane, for arbitrary values of $p$ -- but these solutions will
necessarily be smeared along certain directions\footnote{Further solutions, generated
from certain intersecting configurations of M-branes, were considered
by Emparan in~\cite{emparan:01} but we will say
no more about these here.}.

One of the motivations behind this work was to try to get a handle on
the possible supergravity description of the dielectric branes of
Emparan~\cite{emparan:97} and Myers~\cite{myers:99}.  Imposing T-duality invariance on the non-abelian action
relevant to the description of multiple D$p$--branes, gives rise to a whole
host of worldvolume couplings which are not present in the abelian
theory~\cite{myers:99,taylor:99}.  In particular, the Chern-Simons piece of the action must
include couplings to Ramond-Ramond (R-R) potentials of degree \emph{greater} than
$(p+1)$, giving rise to possible couplings to
higher-dimensional branes.  The presence of such terms allows for the
``dielectric effect'', the simplest example of which is that of $N$
D$p$-branes expanding into a D$(p+2)$-brane under the influence of the
$(p+3)$-form R-R potential associated with the latter~\cite{myers:99,trivedi:00}.  The resulting
dielectric brane, which is a minimum of the worldvolume energy
functional, has worldvolume $\bM^{p+1} \otimes S^2_{\rm NC}$, where
$S^2_{\rm NC}$ is the non-commutative or fuzzy two-sphere.

Although this process is most properly described from within the
non-abelian worldvolume theory of the D$p$-branes, one can
alternatively consider the dual description from within the abelian
theory of the D$(p+2)$-brane.  In the large $N$ limit, the non-commutative nature of the sphere is
lost, and these two descriptions agree~\cite{myers:99}.  Since it is unlikely that such non-commutative
structures will appear within a supergravity perspective, any supergravity
solution purporting to describe this dielectric effect should be
compared to solutions of the D$(p+2)$-brane worldvolume theory.
From this perspective, the standard coupling of a $(p+1)$-form
R-R potential to the D$(p+2)$-brane gives rise to ``dissolved''
D$p$-branes within the worldvolume of the D$(p+2)$-brane~\cite{douglas:95}.  And
it is the presence of this dissolved D$p$-brane charge which allows
for spherical solutions of the D$(p+2)$-brane theory.  

The existence of such spherical D$(p+2)$-branes with dissolved
D$p$-brane charge was anticipated by Emparan in~\cite{emparan:97}.
The main point of this work, however, was to analyse spherical
D$(p+2)$-brane solutions with dissolved F-string charge.  The
interpretation of such solutions was that of $N$ ``Born-Infeld strings''~\cite{callan:97,gibbons:97}
expanding into a sphere under the influence of an external R-R
potential.  Various arguments can then be used to identify the string-like
solution of Born-Infeld theory as a \emph{fundamental} string, but it
is clear that one lacks a proper description of this process from the
point of view of the F-strings themselves.

Many features of the supergravity solutions discovered by Costa
\emph{et al}~\cite{costa:01} and Emparan~\cite{emparan:01} suggest that they
should, indeed, be thought of as the proper gravitational description
of these dielectric branes for $p=4$\footnote{Corresponding, unsmeared
supergravity solutions for different values of $p$ should exist, but
they cannot be generated using the techniques of
~\cite{costa:01,emparan:01}.}.  We
will provide new evidence which significantly strengthens the case for
such a connection, and clarifies the relationship between these geometries
and the unstable solutions of~\cite{dowker:95}.

In sections \ref{sect:myers} and \ref{sect:emparan}
we review possible spherical solutions
of the D6-brane worldvolume theory due to the presence of dissolved
D4-brane and F-string charge respectively.  We comment briefly on the
more general case of the spherical D$p$-brane with dissolved F-string
charge, for arbitrary $p$, the $p=2$ case being qualitatively
different to the other examples.  In section \ref{sect:sugra}, we turn
to the supergravity solutions of~\cite{costa:01,emparan:01}.  To
comment on them, a brief review of their construction is necessary.  We treat
both the F-string and D4-brane solutions in a unified way, since they
have the same structure, and then compare directly with the 
worldvolume analysis of sections \ref{sect:myers} and \ref{sect:emparan}.  
In both cases, the solution is consistent for precisely two values of the radius of the
five- or two-dimensional sphere, a point missed in the treatment of~\cite{costa:01} because
the limiting procedure used therein sends one of the radii to infinity.
Moreover, the functional form of these radii as a function of $N$, the
number of D4-branes or F-strings, precisely matches that of the world volume
calculation.  That this should be the case is rather unexpected.
Moreover, a brief consideration of the smeared solutions of
Emparan~\cite{emparan:01} shows that the atypical structure of the
D2-brane with dissolved F-string charge is also captured by the
corresponding supergravity solution.

In subsection \ref{sect:energy}, we comment on the energetics of these two consistent solutions, and show
that that with the smaller value of the radius 
has the lower energy, in agreement with the worldvolume picture.
We then turn to a consideration of ``off-shell''
configurations, those with arbitrary values of the radius of the
sphere.  In general, such solutions exhibit conical deficits
which can be viewed as ``deficit branes'' providing the tension
necessary to hold the dielectric brane in equilibrium.
We then argue that the tension of these deficit branes gives us a handle on the
stability of the supergravity solutions, with the results that the solution
at the smaller radius is in fact stable -- at least to radial
perturbations -- and that the second solution is
unstable\footnote{We are grateful to Roberto Emparan for pointing out
these arguments to us.}.  These additional pieces of evidence exhibit the
close relationship between the worldvolume and supergravity
descriptions. We believe further study of this connection will greatly
enhance our understanding of the dielectric effect.

\section{Spherical D6-branes from D4-branes -- the worldvolume theory}
\label{sect:myers}

Consider, then, $N$ flat D4-branes polarized by a 7-form R-R potential into
a D6-brane.  From the point of view of the D6-brane, such a
configuration is described by a solution with topology $\bM^5 \otimes
S^2$ and $N$ units of dissolved D4-brane flux.  The flat space
approximation\footnote{``Approximation'' in the sense that it is
not a consistent supergravity background.} analogous to that considered by
Myers~\cite{myers:99} -- a flat background
geometry and a constant 8-form field strength -- does indeed allow for such
solutions.  We will see that the supergravity solutions of Costa \emph{et al}~\cite{costa:01} and
Emparan~\cite{emparan:01} correspond to this simple model with a
surprising degree of accuracy.

The flat ten-dimensional metric is written as
\be
\label{flatmetric}
{\rm d}s^2 = {\rm d}s^2(\bM^5) +{\rm d}r^2 +r^2{\rm d}\Omega^2_2 +{\rm d}s^2(\bE^2),
\ee
where ${\rm d}\Omega^2_n$ denotes the round metric on a unit
$n$-sphere, and the worldvolume of the D6-brane is taken to be the obvious $\bM^5 \otimes
S^2$.  We are interested in static solutions of the worldvolume theory for
$r=R=\mbox{const}$.  The constant background 8-form field strength and
corresponding 7-form potential is chosen to be
\ba
F_{[8]} &=& -2f r^2 \epsilon(\bM^5) \wedge {\rm d}r \wedge
\epsilon(S^2), \\
C_{[7]} &=& \frac{2}{3} fr^3 \epsilon(\bM^5) \wedge \epsilon(S^2),
\ea
where $\epsilon({\cal M})$ denotes the volume form on the space ${\cal
M}$, and where a convenient choice of gauge has been made.

The D6-brane action is a sum of Dirac-Born-Infeld~\cite{leigh:89} and
Chern-Simons~\cite{douglas:95,li:95,green:96} terms:
\ba
\label{wvaction}
S &=& S_{\rm BI} + S_{\rm CS}, \\
S_{\rm BI} &=& -\frac{T_6}{g_s} \int_{\rm D6} e^{-\phi}
\sqrt{-\det\left(P[G]_{ab} + \la F_{ab} \right)}, \\
S_{\rm CS} &=& \frac{T_6}{g_s}\int_{\rm D6} P\left[\sum C_{[n]}\right]
\wedge e^{\la F},
\ea
where we have set the Kalb-Ramond 2-form to zero, $F$ is the
abelian Born-Infeld 2-form field strength, $P[\ldots]$ denotes
a pull-back to the worldvolume of the spacetime fields,
$T_6=\mu_6=((2\pi)^6\al'^{7/2})\inv$ and $\lambda=2\pi \alpha'$.  The Chern-Simons
action is a sum of terms containing all the R-R potentials,
$C_{[n]}$, for $1 \le n \le 7$:
\be
S_{\rm CS} = \frac{T_6}{g_s} \int_{\rm D6} \left( P\left[C_{[7]}\right] +
\la P\left[C_{[5]}\right] \wedge F + \ldots \right),
\label{chern-simons}
\ee
which are the relevant couplings of these potentials to D$p$-branes for $0 \le
p \le 6$.  The Born-Infeld field strength is
then chosen such that the second term in (\ref{chern-simons}) mimics
the coupling of $N$ D4-branes to $C_{[5]}$.  With
\be
F = \frac{N}{2} \epsilon(S^2),
\ee
we have
\be
\lambda T_6 \int_{\mathbb{M}^5 \otimes S^2}
P\left[C_{[5]}\right] \wedge F = N T_4 \int_{\mathbb{M}^5} P\left[C_{[5]}\right],
\ee
where we have used the relation $2\pi \la T_{p+2}=T_{p}$.

\begin{figure}
\center
\epsfig{file=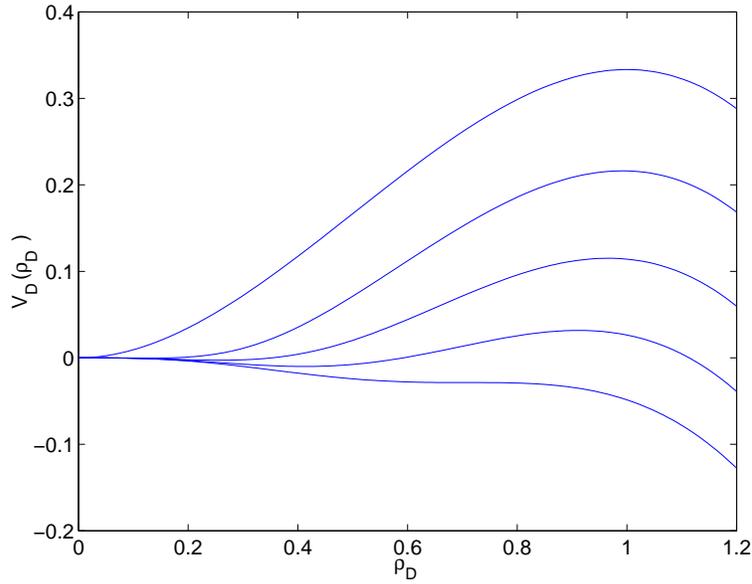,width=10cm}
\flushleft
\caption{This plot shows the potential energy, (\ref{dimless_V}), of
the spherical D6-brane for different values of dissolved D4-brane charge, $N_D$
(\ref{NW}). The top curve is for $N_D=0$ and the lower one
for $N_D=\frac{1}{2}$. The potential has been shifted by a constant
such that the energy at $\rho_D=0$ is zero.
\label{fig:potential}
}
\end{figure}

\begin{figure}
\center
\epsfig{file=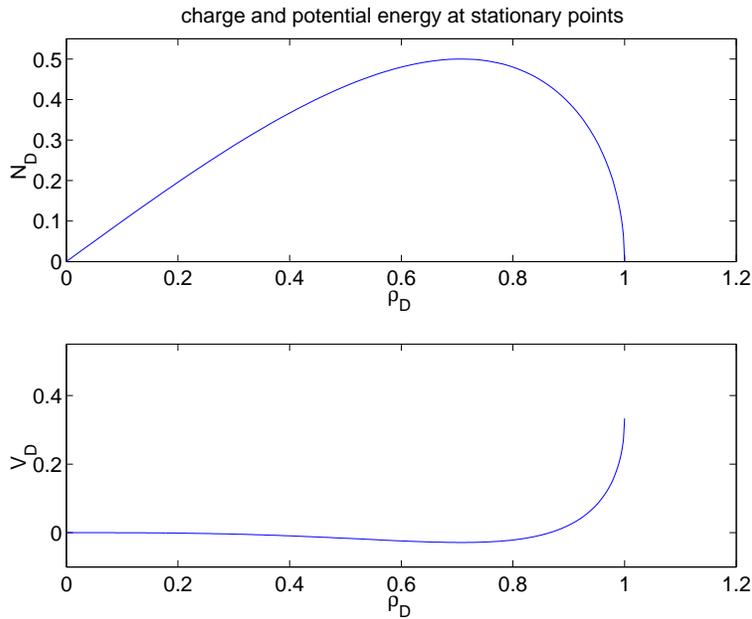,width=10cm}
\flushleft
\caption{Here we show how the charge, $N_D$ (\ref{Nmyers}), 
and energy, ${\cal V}_D (\ref{dimless_V})$,
at the stationary points vary with the radius of the spherical D6-brane
with dissolved D4-brane charge.
\label{fig:myersQE}
}
\end{figure}

Substituting for this Born-Infeld field strength, and for the radius $r=R$, in the background metric
(\ref{flatmetric}), we find that the action (\ref{wvaction}) gives the potential~\cite{myers:99}
\be
\label{wvpotential}
{\cal V}(R) = \frac{4\pi T_6 V_4}{g_s} \left( \sqrt{ R^4+\frac{1}{4}\lambda^2 N^2
} -  \frac{2}{3}fR^3 \right),
\ee
where we have taken the dilaton to vanish, and where $V_n$ denotes the
volume of $\bE^n$.  If we
introduce the dimensionless worldvolume quantities
\ba
\rho_D&=&fR,\\
\label{NW}
N_D&=&\frac{1}{2}f^2\lambda N,\\
\label{dimless_V}
{\cal V}_D&=&\frac{g_s}{4\pi T_6 V_4} f^2 {\cal V} 
      =\sqrt{\rho_D^4+N_D^2}-\frac{2}{3}\rho_D^3,
\ea
then the dimensionless potential, ${\cal V}_D$, has extrema when the following condition holds:
\be
\label{Nmyers}
N_D^2 = \rho_D^2-\rho_D^4.
\ee
For $N_D<\frac{1}{2}$ there are two non trivial extrema, given by
\be
\rho_D^2 = \half \left( 1 \pm \sqrt{1 - 4 N_D^2} \right),
\ee
as shown in figure \ref{fig:potential}.  There are thus two
solutions describing spherical D6-branes, one a maximum of the potential, the other a local
minimum.  In figure \ref{fig:myersQE}a we show how the stationary points of the
potential change as the number of D4-branes change.  Figure
\ref{fig:myersQE}b shows the value of the potential at these extrema.

Myers~\cite{myers:99} considers the case where $N_D<\frac{1}{2}$ in the limit
$N_D\gg\rho_D^2$ in which case one may expand the potential
(\ref{dimless_V}) as ${\cal V}_D\simeq
N_D+\frac{\rho_D^4}{2N_D}-\frac{2}{3}\rho_D^3+\ldots$.  The single non-zero
extremum is then at $\rho_D\simeq N_D$ so, in this limit, we lose the
information about the maximum.  A similar
limit was taken in the supergravity solution of~\cite{costa:01} which again
obscured the fact that there is a second consistent solution.
We shall see below that, prior to any limit being taken, the allowed
radii of the supergravity solution have precisely
the same functional form as (\ref{Nmyers}).  We believe this to be strong evidence that
the worldvolume and supergravity pictures are describing the same phenomenon.

\section{Spherical D6-branes from F-strings -- the worldvolume theory}
\label{sect:emparan}

\begin{figure}
\center
\epsfig{file=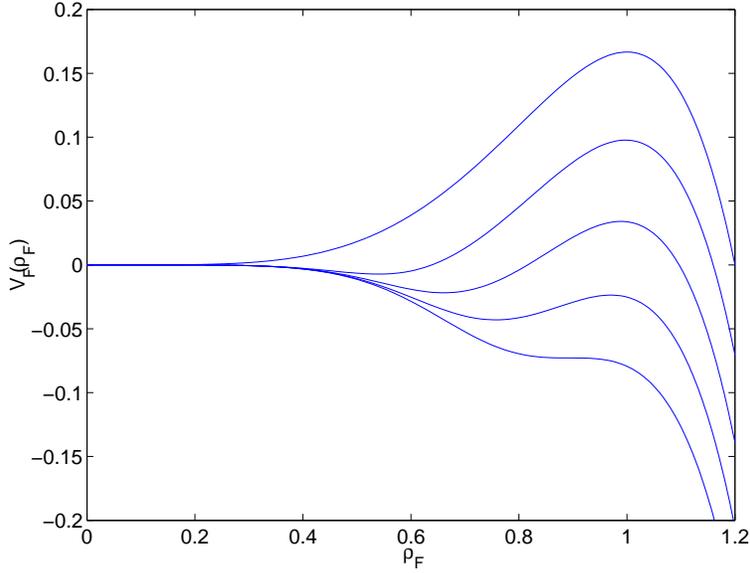,width=10cm}
\flushleft
\caption{This plot shows the potential energy, (\ref{dimless_Vf}), of
 the spherical D6 brane for different values of dissolved F-string
 charge, $N_F$ (\ref{dimless_N}).  The top curve
 is for $N_F=0$ and the lower one for the maximum charge
 $N_F=\sqrt{\left(\frac{4}{5}\right)^4-\left(\frac{4}{5}\right)^5}$.
 The potential has again been shifted by a constant to ensure that the zero
 of energy is at $\rho_F=0$.
\label{fig:fpotential}
}
\end{figure}

\begin{figure}
\center
\epsfig{file=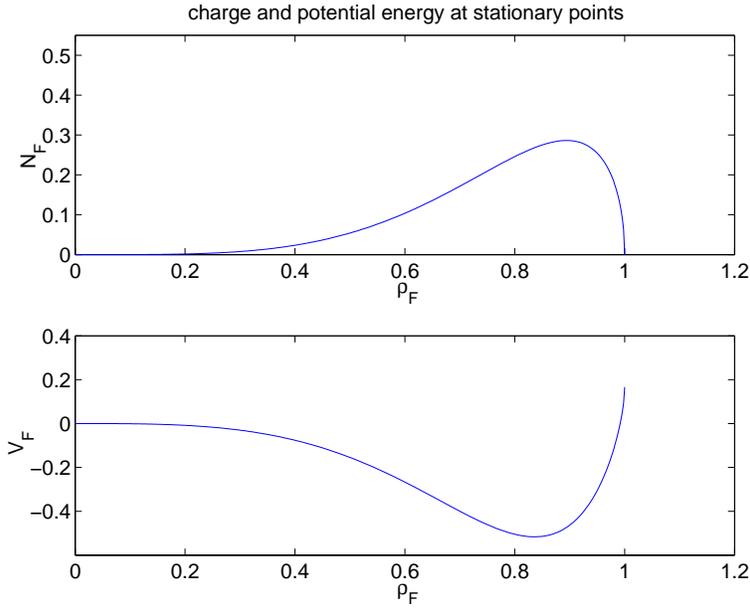,width=10cm}
\flushleft
\caption{Here we show how the charge, $N_F$ (\ref{Nf}), 
and energy, ${\cal V}_F (\ref{dimless_Vf})$,
at the stationary points vary with the radius of the spherical D6-brane with dissolved F-string charge.
\label{fig:emparanQE}
}
\end{figure}

We now turn to the case considered by Emparan~\cite{emparan:97},
in which $N$ F-strings expand into a D6-brane with topology
$\bM^2 \otimes S^5$.  Taking the same action as in (\ref{wvaction}), 
we place the D6-brane at constant $r=R$
in the flat spacetime given by
\be
{\rm d}s^2= {\rm d}s^2(\bM^2) +{\rm d}r^2 +r^2{\rm d}\Omega^2_5 +{\rm d}s^2(\bE^2).
\ee
The Born-Infeld field strength is taken to be
\be
F=\E \epsilon (\bM^2),
\ee
where $\E$ is a constant electric field, and we take the constant R-R
8-form field strength and corresponding 7-form potential to be
\ba 
F_{[8]}&=&5h \epsilon (\bM^2) \wedge{\rm d}r\wedge 
          r^5\epsilon(\Omega_5),\\
C_{[7]}&=&\frac{5}{6}h \epsilon (\bM^2) \wedge r^6\epsilon(\Omega_5).
\ea
The action (\ref{wvaction}) then reduces to
\be
S=-\frac{T_6\Omega_5l}{g_s}\int{\rm d}t
   \left(R^5\sqrt{1-\lambda^2 \E^2}-\frac{5h}{6}R^6\right) \equiv \int{\rm
d}t L,
\ee
where $l$ is the length of the string and $\Omega_n$ denotes the
volume of a unit $n$-sphere.  Introducing the conjugate
momentum, $\D=\frac{\delta L}{\delta \E}$, we find that the
Hamiltonian, $(\E \D-L)$, is given by~\cite{emparan:97}
\be
H=\frac{T_6\Omega_5l}{g_s}
\left(\sqrt{R^{10}+\left(\frac{g_s}{T_6\Omega_5\lambda}\right)^2\D^2}-\frac{5h}{6}R^6\right).
\ee
To relate $\D$ to $N$, the number of dissolved F-strings, we take $R$ to zero
to obtain a string-like state with energy per unit length $\D/\lambda=\D T_{\rm F}$
($T_{\rm F}$ being the F-string tension), so we
simply associate $\D=N$~\cite{emparan:97}.  Now we use 
$R_{11}=g_s\sqrt{\alpha'}$, $\Omega_5=\pi^3$, and
introduce the following dimensionless quantities
\ba
\label{dimless_rho}
\rho_F&=&hR,\\
\label{dimless_N}
N_F&=&8h^5 R_{11}\lambda^2N,\\
\label{dimless_Vf}
{\cal V}_F&=&8\lambda^3R_{11}h^5 \frac{H}{V_1} = \sqrt{\rho^{10}_F+N^2_F}-\frac{5}{6}\rho^6_F.
\ea
The extrema of the potential (\ref{dimless_Vf}) are given by solutions of
\be
\label{Nf}
N_F^2=\rho_F^8-\rho^{10}_F.
\ee
Again there is a maximum charge, above which there are no stable solutions. To
calculate it, we use the fact that at this maximum charge, the two extrema
merge into a point of inflection, so ${\cal V}_F'={\cal V}_F''=0$ at the
extrema.  This gives $N_F\leq\sqrt{\left(\frac{4}{5}\right)^4-\left(\frac{4}{5}\right)^5}$.
In figure \ref{fig:fpotential} we show how the potential changes with the
number of F-strings dissolved in the D6-brane.
How the location of the extrema changes with $N_F$ is
shown in figure \ref{fig:emparanQE}a, there clearly being two extrema
for $N_F$ less than the maximum value.  This system is seen to be
qualitatively the same as the one considered in section
\ref{sect:myers}, the D4-brane blowing up into a D6-brane.

\subsection{Spherical D$p$-branes from F-strings}
\label{sect:f1_d2}

The above analysis is not peculiar to D6-branes, in that spherical
D$p$-branes with dissolved F-string charge exist for arbitrary values
of $2\le p \le 8$~\cite{emparan:97}.  It is easy to show that the
dimensionless potential in the general case has the form~\cite{emparan:97}
\be
{\cal V}_F = \sqrt{\rho^{2(p-1)}_F+N^2_F}-\frac{p-1}{p}\rho^p_F,
\ee
in terms of dimensionless quantities analogous to those defined in
(\ref{dimless_rho}) and (\ref{Nf}) above.  The extrema of this
potential are given by solutions of
\be
\label{eqn:wvol_f1_dp}
N_F^2=\rho_F^{2(p-2)}-\rho_F^{2(p-1)},
\ee
so for $p>2$ there are always two extrema.  The $p=2$ case is
qualitatively different, however.  There is no minimum in this case,
the single non-trivial extremum being at
\be
\label{eqn:max_d2}
\rho_F=\sqrt{1-N_F^2},
\ee
and this is a maximum of the potential.  We will see that the relevant
(smeared) supergravity solution describing this configuration has a similar structure.

\section{The supergravity description}
\label{sect:sugra}

The eleven-dimensional starting point of the discussion of~\cite{costa:01,emparan:01} is
the double analytic continuation of the rotating black M2- and
M5-branes\footnote{The standard versions of which are to be found in~\cite{cvetic:96,csaki:99}.}:
\ba
\label{eqn:11d_metric}
{\rm d}s^2&=&
H^{-2/d} \left[ {\rm d}s^2 ( \bM^{d-1} ) + f \left( {\rm
d}\tau + \frac{kl \cosh{\alpha} }{\Delta r^{\td{d}} f}
\sin^2{\theta}{\rm d}\tilde{\phi}\right)^2\right]\\
\nonumber
&& \qquad \qquad +H^{2/\td{d}}\left[\frac{{\rm d}r^2}{\tilde{f}}+
r^2\left(\Delta{\rm d}\theta^2+\frac{\Delta\tilde{f}}{f}\sin^2{\theta}{\rm d}\tilde{\phi}^2
+\cos^2{\theta}{\rm d}\Omega_{\td{d}-1}^2\right)\right],
\ea
where $(d,\td{d})=(3,6)$ and $(6,3)$ for the M2- and M5-branes
respectively, and where
\be
\label{eqn:functions}
H=1+\frac{k\sinh^2{\alpha}}{\Delta r^{\tilde{d}}}, \qquad
f=1-\frac{k}{\Delta r^{\tilde{d}}}, \quad
\tilde{f}=\frac{r^{\tilde{d}}-l^2r^{\tilde{d}-2}-k}{\Delta r^{\tilde{d}}},
\qquad \Delta=1-\frac{l^2\cos^2{\theta}}{r^2}.
\ee
For the M2-brane, the 3-form potential is given by
\be
\label{eqn:3form_m2}
A_{[3]} = \frac{k\sinh{\alpha}}{\Delta r^{\td{d}} H} \ep (\bM^2) \wedge
\left[ \cosh{\alpha} {\rm d}\t - l \sin^2{\th} {\rm d} \td{\p} \right],
\ee
whereas for the M5-brane it is
\be
\label{eqn:3form_m5}
A_{[3]} = \frac{k\sinh{\alpha}}{\Delta}\cos^3{\theta}
\left[-\left(1-\frac{l^2}{r^2}\right)\cosh{\alpha}
{\rm d}\tilde{\phi}-\frac{l}{r^2}{\rm
d}\tau\right]\wedge\epsilon\left(S^2\right).
\ee
The location, $r_H$, of the ``Euclidean horizon'' or bolt is given by
the zero of $\tilde{f}$, so that
\be
\label{horizon}
k = r_H^{\tilde{d}-2}\left(r_H^2-l^2\right),
\ee
and the ``Euclidean angular velocity'', $\Omega$, is
\be
\Omega = \frac{l}{\cosh{\alpha}\left(r_H^2-l^2\right)}.
\ee
To avoid a conical singularity at $r=r_H$, $\tau$ must be periodic with
period $2\pi R_{11}$, where
\be
\label{radius11}
R_{11} = \frac{2 k \cosh{\alpha}}
{\tilde{d}r_H^{\tilde{d}-1}-(\tilde{d}-2)l^2r_H^{\tilde{d}-3}} = g_s\sqrt{\al'}.
\ee
Finally, we must also consider the quantization of M2- and M5-brane
charge which gives
\be
\label{charge}
k\cosh{\alpha}\sinh{\alpha} = c_{\td{d}} N (R_{11} \lambda)^{\td{d}/3}, 
\ee
where we have used the fact that the eleven-dimensional Planck length,
$l_P$, is given by $l_P = g_s^{1/3} \sqrt{\al'}$.  $N$ denotes the
number of M2- or M5-branes, and the constant $c_{\td{d}}$ is
\be
c_{\td{d}} = \frac{(2\pi)^{2\td{d}/3}}{\td{d} \Omega_{\td{d}+1}},
\ee
so that $c_6 = 8$ and $c_3=1/2$.

As explained in~\cite{costa:01,emparan:01}, the ten-dimensional geometry is found
by reducing along orbits of the Killing vector
\be
K = \frac{\partial}{\partial\tau}
    +B\frac{\partial}{\partial\tilde{\phi}},
\ee
the fixed point set of the corresponding isometry being $\{r=r_H$,
$\th=0\}$.  To dimensionally reduce, one identifies points separated by a distance of $2\pi
R_{11}$ along integral curves of $K$.  Introducing a new angular coordinate $\p = \td{\p} - B \t$, which has
standard $2\pi$ periodicity and is constant along orbits of $K$, gives
$K=\del/\del \t$.  Then the ten-dimensional string frame
metric is
\ba
\label{eqn:10d_metric}
{\rm d}s^2&=&
\Sigma^{1/2} H^{-3/d} {\rm d} s^2 ( \bM^{d-1} ) +
\Sigma^{1/2} H^{1-3/d} \left[ \frac{ {\rm d} r^2}{ \tilde{f} } + r^2
\left( \Delta {\rm d} \theta^2 + \cos^2{\theta} {\rm d} \Omega_{\tilde{d}-1}^2\right)\right]\\
\nonumber
&& \qquad\qquad + \Sigma^{-1/2} H^{1-3/d} \tilde{f} \Delta r^2
\sin^2{\theta} {\rm d} \phi^2,
\ea
where
\be
\Sigma = f \left( 1 + \frac{ Bkl \cosh{\al} \sin^2{\th} }{ r^{\td{d}}
\Delta f } \right)^2 + H \frac{\Delta\td{f}}{f}(Br\sin{\th})^2,
\ee 
and the dilaton is given by\footnote{We will not write down the dimensional reduction of
the 3-form potentials (\ref{eqn:3form_m2}) and (\ref{eqn:3form_m5}) since they are
superfluous to our discussion.}
\be
e^{2\phi} = \Sigma^{3/2} H^{-3/d}.
\ee
In general, the metric (\ref{eqn:10d_metric}) has a conical
singularity in the $r$--$\p$ plane at $r=r_H$.  An analysis of the $r \ra r_H$
limit shows that the deficit angle is given by $2\pi(1-a)$, where
\be
\label{eqn:deficit}
a=\half \frac{\td{d}r_H -
(\td{d}-2)l^2r_H\inv}{l+B\cosh{\al}(r_H^2-l^2)}.
\ee
Then by imposing
\be
\label{10Dcone}
B = \frac{1}{R_{11}}-\Omega=\frac{\td{d}r_H+(\td{d}-2)l}{2r_H(r_H+l)\cosh{\alpha}},
\ee
we have $a=1$, and so no conical singularity.  As described in~\cite{costa:01,emparan:01}, the
ten-dimensional metric has the correct form to describe an F-string or
D4-brane expanding into a D6-brane.  The fixed point set
$\{r=r_H,\th=0\}$ of $K$ leads to a null singularity in the metric
(\ref{eqn:10d_metric}), the singular surface being identified with the
spherical part of the worldvolume of the
F-string or D4-brane: $\bM^2 \otimes S^5$ or $\bM^5 \otimes S^2$
respectively.  Moreover, an analysis of the near-core,
$\{r=r_H,\th=0\}$, spacetime shows
that, in both cases, the radius of the sphere into which the F-string
or D4-brane expands is precisely equal to $r_H$~\cite{costa:01,emparan:01}.

\subsection{Smeared solutions: F-strings expanding into a D$p$-brane}
\label{sect:sugra_f1_dp}

We have seen in section \ref{sect:f1_d2}, from the worldvolume point
of view, that the case of F-strings expanding into a D2-brane is qualitatively different to
that of expansion into any other D$p$-brane: there is only one static solution and it is unstable.
Here we obtain a supergravity description, albeit a smeared 
one~\cite{emparan:01}, describing $N$ F-strings expanding into a
D$p$-brane, for arbitrary $2\le p<6$. We will see
that the $p=2$ is a special case in supergravity as well.
The starting point
in this case is a smeared version of the M2-brane metric (\ref{eqn:11d_metric}):
\ba
{\rm d}s^2&=&
H^{-2/3} \left[ {\rm d}s^2 ( \bM^2 ) + f \left( {\rm
d}\tau + \frac{kl \cosh{\alpha} }{\Delta r^p f}
\sin^2{\theta}{\rm d}\tilde{\phi}\right)^2\right]\\
\nonumber
&& \qquad \qquad +H^{1/3}\left[{\rm d}s^2(\bE^{6-p}) + \frac{{\rm d}r^2}{\tilde{f}}+
r^2\left(\Delta{\rm d}\theta^2+\frac{\Delta\tilde{f}}{f}\sin^2{\theta}{\rm d}\tilde{\phi}^2
+\cos^2{\theta}{\rm d}\Omega_{p-1}^2\right)\right],
\ea
where the functions $H,f,\td{f}$ and $\Delta$ are given by
(\ref{eqn:functions}), but with $\td{d}$ replaced by $p$.  The
solution is thus smeared over the obvious $6-p$ transverse
directions.  The relations (\ref{horizon}) and (\ref{radius11}) are
unchanged up to the replacement of $\td{d}$ with $p$, as is the 3-form
potential (\ref{eqn:3form_m2}).  The charge quantization
condition is then
\be
k\cosh{\alpha}\sinh{\alpha} = c_p N \frac{(R_{11}
\lambda)^2}{V_{6-p}},
\ee
where
\be
c_p = \frac{(2\pi)^4}{p \Om_{1+p}}.
\ee
Dimensional reduction proceeds as above, and gives rise to a D6-brane
with topology $\bM^2\times \bE^{6-p} \times S^{p-1}$, so that T-duality
along the flat transverse directions gives the desired solution
describing $N$ F-strings expanding into a D$p$-brane with topology
$\bM^2 \times S^{p-1}$~\cite{emparan:01}.  Again, the potential
conical singularity is avoided if $B$ is given by (\ref{10Dcone}), up
to the replacement of $\td{d}$ with $p$.

\subsection{The radius of the sphere}

The question with which we are concerned here is, for what values of
$r_H$ does the solution (\ref{eqn:10d_metric}) make sense?  
That is, what values of $r_H$
are allowed and, furthermore, do the allowed
values of $r_H$ match the radii of the polarized branes
discussed in sections \ref{sect:myers} and \ref{sect:emparan}?
To address this question, we need to
think about which quantities to hold fixed as we vary the values of
$N$ and $B$.  At first sight, one might consider holding the rotation
parameter, $l$, fixed, but this is incorrect; we should rather be
considering the set of possible solutions for constant $2
\ka_{10}^2=(2\pi)^4 R_{11}^2\la^3$.  In other words, we should fix
$R_{11}$, in order that our family of ten-dimensional solutions have the same Newton's
constant.

\begin{figure}
\center
\epsfig{file=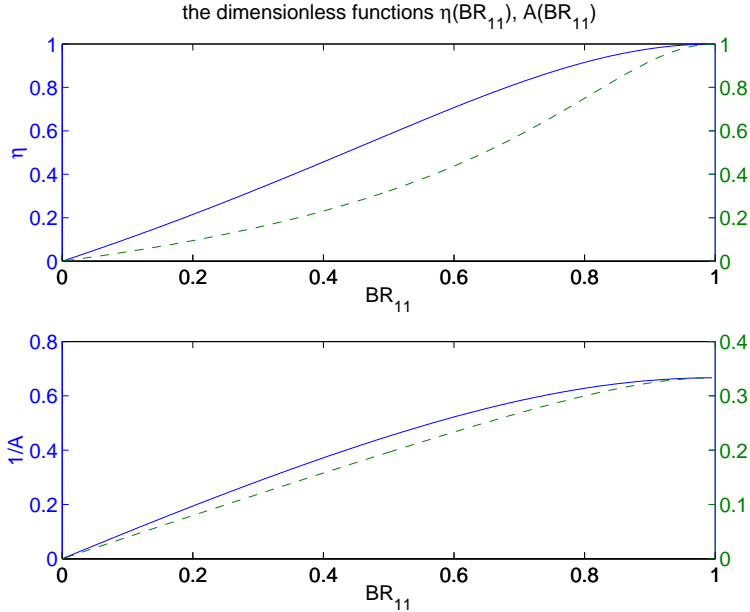,width=10cm}
\flushleft
\caption{In order to find $r_H$ for a given set of parameters,
$R_{11}$, $B$ and $N$ we need the dimensionless functions $A(BR_{11})$,
$\eta(BR_{11})$.  (We plot $A^{-1}(BR_{11})$ for clarity.)  The scales
on the left of the graphs, and the solid lines, correspond to $\tilde{d}=3$ (D4$\rightarrow$D6)
and the right hand scales, and dashed lines, correspond to $\tilde{d}=6$ (F1$\rightarrow$D6).
\label{fig:abr11}
}
\end{figure}

We thus use the four relations (\ref{horizon}), (\ref{radius11}),
(\ref{charge}) and (\ref{10Dcone}) to eliminate $l$,
$\alpha$ and $k$, and to give a relation between $r_H$, $N$, $B$ and
$R_{11}$.  First rewrite (\ref{10Dcone}) as
\be
\label{cone}
lr_H^{\tilde{d}-1}-l^2r_H^{\tilde{d}-2}+kB\cosh{\alpha}r_H-\frac{\tilde{d}}{2}k=0,
\ee
Now take (\ref{horizon}) to eliminate $l$ from (\ref{cone}) and
(\ref{radius11}). Then eliminate $\alpha$ by using (\ref{radius11}), giving
\be
\label{eqn:k_eqn}
\frac{1}{4}(\tilde{d}-2)^2(1-BR_{11})^2k^2
+\left[(\tilde{d}-2)(1-BR_{11})^2+1\right]r_H^{\tilde{d}}k
+\left[(1-BR_{11})^2-1\right]r_H^{2\tilde{d}}=0.
\ee
We may now solve this quadratic for $k$ and substitute into (\ref{charge}).
We note that the requirement $k \ge 0$ gives the restriction
\be
0\le BR_{11} \le 1,
\ee
so that $B=1/R_{11}$ is the maximum magnetic
field~\cite{dowker:95}.  However, for the ten-dimensional Kaluza-Klein picture to be valid,
we need all ten-dimensional length
scales to be larger than the compactification scale, $R_{11}$.  The
magnetic field gives just such a length scale, $1/B$, so the solution
is truly ten-dimensional only for $BR_{11}\ll 1$.  At any rate, we now find
\be
k = \eta r_H^{\td{d}},
\ee
where we have defined
\be
\eta =\frac{2}{(\tilde{d}-2)^2(1-BR_{11})^2}
       \left[\sqrt{1+\tilde{d}(\tilde{d}-2)(1-BR_{11})^2}-(\tilde{d}-2)(1-BR_{11})^2-1\right].
\ee
With
\be
A=\frac{1}{\eta}\left(1+\frac{\tilde{d}-2}{2}\eta\right),
\ee
we have
\be
\eta^2c^2_{\tilde{d}}\frac{1}{A^{2\tilde{d}}}
    \left(\frac{\lambda}{R_{11}^2}\right)^{\frac{2\tilde{d}}{3}}N^2
   +\left(\frac{r_H}{AR_{11}}\right)^{2\tilde{d}-2}
   -\left(\frac{r_H}{AR_{11}}\right)^{2\tilde{d}-4} = 0.
\ee
The functions $A(BR_{11})$ and $\eta(BR_{11})$
which control this equation, are plotted in figure
\ref{fig:abr11}.  By introducing the dimensionless quantities
\ba
\label{r-rS}
\rho_S&=&\frac{r_H}{A R_{11}},\\
\label{N-NS}
N_S&=&\frac{c_{\tilde{d}}}{\eta A^{\tilde{d}}}
      \left(\frac{\lambda}{R_{11}^2}\right)^{\frac{\tilde{d}}{3}}N,
\ea
relevant for the supergravity solutions, we finally find
\be
\label{NS}
N_S^2=\rho_S^{2(\tilde{d}-2)}-\rho_S^{2(\tilde{d}-1)},
\ee
which is of precisely the same form as
(\ref{Nmyers}) for $\td{d}=3$, and (\ref{Nf}) for $\td{d}=6$.
Although we cannot reproduce the correct numerical factors for any
value of $BR_{11}$, it is surprising enough that the
form of the equations are the same when one considers that the
worldvolume calculation did not make use of a consistent supergravity background.

In figure \ref{fig:SUGRAQE} we show how the radius, $\rho_S$, of the
$(\td{d}-1)$-sphere depends on $N_S$, the dimensionless quantity
defining the $(d-2)$-brane charge of the solution.  There are various
noteworthy points.  For $N_S$ less than some critical value, there are
always \emph{two} values of $r_H$, which we denote by $r_+$ and $r_-$ (where $r_+ > r_-$).
As long as the F-string or D4-brane charge is not too large, then, there are two distinct
supergravity solutions of the form (\ref{eqn:10d_metric}), for which the spacetime has no conical
singularities.  In the following subsection, we will see that the solution with 
$r_H=r_-$ has the lower energy of the two.  
As we decrease the charge,  
$r_+$ increases, and $r_-$ decreases.  For zero
charge -- essentially the situation considered by Dowker \emph{et
al}~\cite{dowker:95} -- there is only a single non-trivial value of $r_H$,
this solution being unstable~\cite{dowker:95}.  It
corresponds to $\al=0$, whereas the trivial solution $r_H=0$
corresponds to $k=0$, which we shall use as a background to calculate the
energy of the charged solutions.

We now have all the information we need to consider the $B\rightarrow 0$
limit for fixed $N$ and $R_{11}$ (we consider the case $\tilde{d}=3$).
For $BR_{11}\ll 1$, $\eta\propto BR_{11}$ and $A^{-1}\propto BR_{11}$
(see figure \ref{fig:abr11}), so from (\ref{N-NS}) we have $N_S\rightarrow A^{-2}\rightarrow 0$.
Figure \ref{fig:SUGRAQE} shows that there are two radii for which 
$N_S\rightarrow 0$, one at $\rho_S\rightarrow 0$ and the other at
$\rho_S\rightarrow 1$. The former has $N_S\propto\rho_S$ 
(see figure \ref{fig:SUGRAQE}) as $\rho_S\rightarrow 0$
so, from (\ref{r-rS}), we have $\frac{r_H}{R_{11}}\propto\frac{1}{A}\rightarrow 0$.
The latter has $\rho_S\rightarrow 1$ so 
$\frac{r_H}{R_{11}}\propto A \rightarrow \infty$.  The solution with 
$r_H=r_-$ was identified as the stable solution of the worldvolume
theory, so as the magnetic field is reduced this radius goes to
zero; the effect disappears, as expected.  The other solution, 
$r_H=r_+$, is a charged version of the one found
by Dowker \emph{et al}~\cite{dowker:95} where it was shown that the small
$B$ limit gave $r_H\rightarrow\infty$, an effectively planar brane.

It is not difficult to generalize the above analysis to include the
smeared solution describing $N$ F-strings expanding into a D$p$-brane for
arbitrary $2\le p<6$, as discussed in section \ref{sect:sugra_f1_dp}.  For
$p\ne2$, we need only everywhere replace $\td{d}$ with $p$.  The
ten-dimensional solutions thus obey
\be
N_S^2=\rho_S^{2(p-2)}-\rho_S^{2(p-1)},
\ee
which is of the same form as (\ref{eqn:wvol_f1_dp}).  The $p=2$ case
is simpler since the equation analogous to
(\ref{eqn:k_eqn}) is then only linear in $k$.  We have
\be
k=\left[1-(1-BR_{11})^2\right] r_H^2,
\ee
so that, in terms of the relevant dimensionless quantities, we find
\be
\rho_S=\sqrt{1-N_S^2},
\ee
which is of the same form as (\ref{eqn:max_d2}).  We note that
this single consistent solution corresponds to the \emph{larger}
of the two radii, $r_H=r_+$, in the case of general $p$.  As we will see below, it is
only the \emph{smaller} radius which is stable, so it would seem that
F-strings cannot expand into a stable spherical D2-brane.  At any rate, the
atypical properties of the spherical D2-brane with dissolved F-string
charge are precisely captured by the corresponding (smeared)
supergravity solution.  We expect that a localized solution should exist,
and that it should also exhibit the same behaviour.

\subsection{Energetics and stability}
\label{sect:energy}

Following~\cite{costa:01}, to compute the energy of the
ten-dimensional solutions we make use of the background
subtraction method~\cite{hawking:95}, where the relevant background is the
$k=0$ solution: the standard Melvin-like flux 7-brane solution~\cite{costa:00} written in spherical
oblate coordinates.  That is to say we define the zero of energy
to be the $k=0$ solution.  
We also need to transform the metric (\ref{eqn:10d_metric}) to the Einstein
frame for this calculation, using ${\rm d}s^2_{\rm E}=e^{-\p/2}{\rm d}s^2_{\rm S}$.
Specifically~\cite{hawking:95}
\be
\label{eqn:energy}
\E = -\frac{1}{2 \ka_{10}^2} \left[\int_{\infty} d^8 x \sqrt{h}
{\cal N} {\cal K} - \int_{\infty} d^8 x \sqrt{h_0}
{\cal N}_0 {\cal K}_0 \right],
\ee
where $h_{ij}$ is the metric induced on a constant $r$ slice of the
constant time slice, $h$ is its determinant
and the integration is performed as $r\rightarrow\infty$.  The lapse
function, ${\cal N}$, and extrinsic curvature, ${\cal K}$, of the
constant $r$ slice, are given by
\be
\label{eqn:N+K}
{\cal N} = \sqrt{-g_{tt}}, \qquad {\cal K} =
\frac{1}{\sqrt{g_{rr}}} h^{ij} \del_r h_{ij},
\ee
and a ``$0$'' subscript denotes the corresponding quantities for the
reference $k=0$ background.

\begin{figure}
\center
\epsfig{file=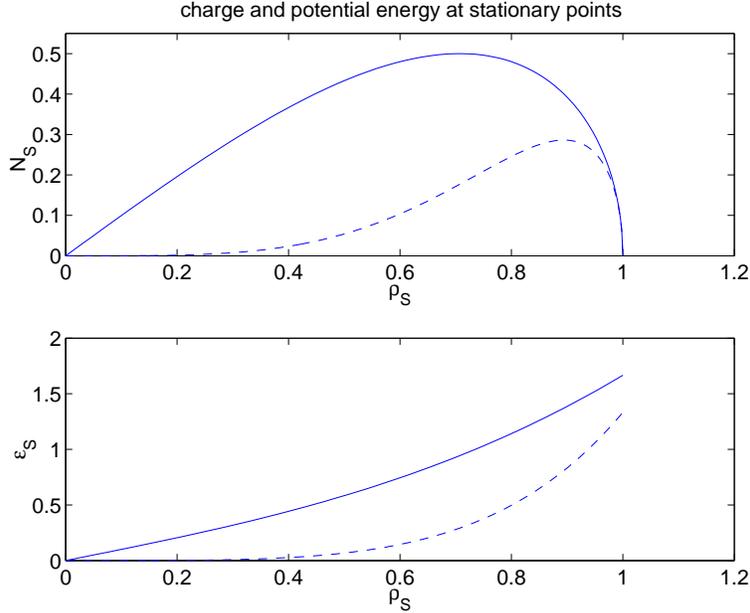,width=10cm}
\flushleft
\caption{Analogous plots to figures \ref{fig:myersQE} and \ref{fig:emparanQE}
for the supergravity case without conical singularities. They
show how the charge, $N_S$ (\ref{NS}), and energy, $\E_S$ (\ref{ES}),
vary as the radii of the static 
solutions change. The solid line is
for $\tilde{d}=3$ (D4$\rightarrow$D6) and the dashed line is for
$\tilde{d}=6$ (F1$\rightarrow$D6).
\label{fig:SUGRAQE}
}
\end{figure}

We find\footnote{The energy of these dielectric branes is thus slightly higher than the
corresponding energy of the standard black F-string and D4-brane solutions, which have
a factor of $(\td{d} + 1)$ rather than $(\td{d} + 2)$~\cite{duff:96}.}
\be
\E = \frac{\Omega_{\tilde{d}+1}V_{d-2}}{(2\pi)^4 R_{11}^2\lambda^3}
     \left(\tilde{d}\sinh^2{\alpha}+\td{d} + 2 \right)k.
\ee
Defining the dimensionless energy, $\E_S$, as
\be
\E_S = \frac{ c_{\tilde{d}} (2\pi)^{2(2-\td{d}/3)} \lambda^3}
            {A^{\tilde{d}}\eta V_{d-2} R_{11}^{\tilde{d}-2}}\E,
\ee
gives
\be
\label{ES}
\E_S = \rho_S^{\tilde{d}-2}\left(1+\frac{2}{\tilde{d}}\rho_S^2\right),
\ee
which is plotted in figure \ref{fig:SUGRAQE} for the two values of
$\td{d}$.  In both cases, the solution with $r_H=r_-$ has lower energy
than that with $r_H=r_+$,
which agrees with the pictures presented in sections \ref{sect:myers}
and \ref{sect:emparan}.

Although the solution with $r_H=r_-$ has lower energy than that with $r_H=r_+$,
this does not imply stability.  We can argue, however, that it
is in fact stable to radial perturbations by looking at ``off--shell''
configurations, that is by holding $B$ and $N$ constant and varying $r_H$.
This still yields a supergravity solution but in general there will be
a conical singularity on the surface of, and inside, the dielectric sphere.
The interpretation of this singularity is that it
provides the tension necessary to hold the dielectric sphere at that radius.
One can associate a ``deficit brane'' with this conical singularity and, just
as for a cosmic string~\cite{vilenkin+shellard}, its tension
is proportional to
the deficit angle, $2\pi(1-a)$, with $a$ as in (\ref{eqn:deficit}).
So if the deficit angle is positive, then the deficit brane has positive
tension, and is having to pull on the dielectric sphere to keep it
static; the dielectric sphere wants to expand.  Similarly, when the
tension is negative, the sphere wants to contract.

To compute the tension (mass per unit volume)
of the deficit brane, we take the
$r\ra r_H$ limit of the metric (\ref{eqn:10d_metric}) and again make use of the
background subtraction method~\cite{hawking:95}, as applied to the
resulting spacetime.  As explained in~\cite{hawking:95}, the relevant
background in this case is a spacetime identical in all respects
except that it is free of the conical singularity.  Denoting the energy in which we are interested by $\T$, and applying
the formulae (\ref{eqn:energy}) and (\ref{eqn:N+K}) to the case at
hand, we find
\be
\label{eqn:tension}
\T = \frac{\Omega_{\tilde{d}+1}V_{d-2}}{(2\pi)^4 R_{11}^2\lambda^3}
r_H^{\td{d}-1} \left( l + B\cosh{\al} (r_H^2-l^2) \right) (1-a)
=\frac{\Omega_{\tilde{d}+1}V_{d-2} R_{11}^{\tilde{d}-2}}{(2\pi)^4\lambda^3}\T_S
\ee
which, as promised, is proportional to the deficit angle.  
$\T_S$ is the mass of the deficit brane in terms of the dimensionless
quantities.  It is interesting to note that we can derive this expression using
different techniques.  In particular, we have considered methods
analogous to those used in~\cite{costa:00b} to compute the 
effective energy-momentum tensor of the conical singularity
between two collinear black holes in four dimensions; and these methods
give exactly the same result for the mass $\T$.

\begin{figure}
\center
\epsfig{file=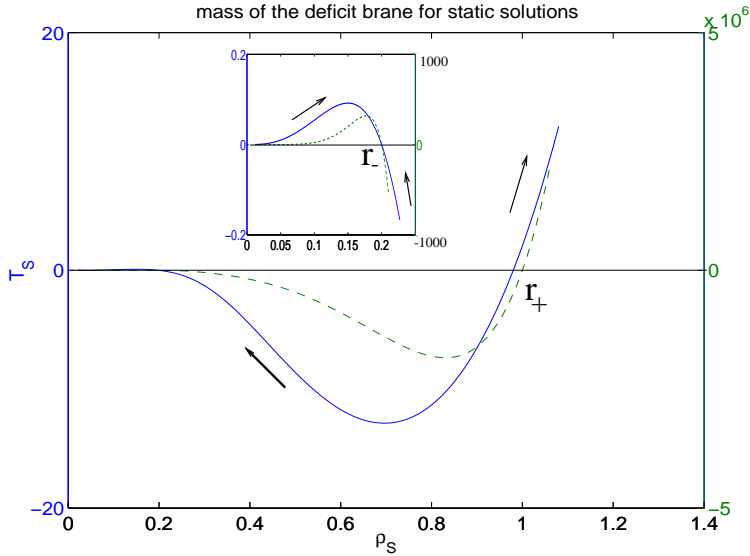,width=10cm}
\flushleft
\caption{This plot shows the dimensionless mass (\ref{eqn:tension}) of
the deficit brane for $\td{d}=3$ (solid line, left hand scale),
$\td{d}=6$ (dashed line, right hand scale), and for some
specific values of $B$ and $N$.  The inset is a magnification of the
region around $r_H=r_-$.  The arrows show whether the
sphere wants to expand or contract as one moves away from the zeros of
$\T_S$.
\label{fig:tension}
}
\end{figure}

To analyse the form of (\ref{eqn:tension}) further, we fix $B$ and
$N$, and substitute for $l$, $\al$ and $k$ using the expressions
(\ref{horizon}), (\ref{radius11}) and (\ref{charge}).  Of course, we
no longer have access to the expression (\ref{10Dcone}) for $B$, since
this was derived to ensure the absence of conical singularities.
Indeed, we can no longer solve for $\al$ or $k$ analytically, and must proceed using numerical
methods. The plot of the mass, $\T_S$, as a function of
$\rho_S$ is shown in figure \ref{fig:tension}, for both the
F-string and D4-brane solutions.

As argued previously, a deficit brane with positive tension indicates
that the dielectric sphere wants to expand, and negative tension
implies that it wants to contract.  We see from figure \ref{fig:tension},
then, that at $r_H=r_+$ the sphere is unstable to both expansion
and contraction, whereas at $r_H=r_-$ the sphere is stable.  We should note that
these arguments apply only to radial perturbations.

\section{Discussion}

We have provided strong evidence that the supergravity solutions
of Costa \emph{et al}~\cite{costa:01} and Emparan~\cite{emparan:01}
are describing the dielectric effect whereby F-strings or D4-branes
expand into spherical D6-branes.  
We have presented the two supergravity solutions 
in a unified manner showing that, in both cases, there are two distinct static solutions
which are free of conical singularities.  Moreover,
the radius of these solutions has precisely the same functional form
as that implied by the worldvolume analyses of
Emparan~\cite{emparan:97} and Myers~\cite{myers:99}.

On the supergravity side we have seen that the energy of the $r_H=r_-$ solution
was lower than that with $r_H=r_+$, 
although this alone does not guarantee stability.  To address the latter issue,
we considered the tension of the deficit brane formed by static
solutions away from $r_H=r_\pm$, which indicated that the $r_H=r_-$ solution was
stable and that with $r_H=r_+$ was unstable.  Thus we see that the supergravity solutions capture
the worldvolume properties with remarkable accuracy.

To further strengthen the interpretation of the supergravity solutions, we
considered the special case of $N$ F-strings expanding into a
D2-brane, for which the worldvolume
analysis indicates that no stable dielectric solution should exist.  Although the supergravity
solution for this configuration was smeared in certain transverse
directions, we have seen that indeed there was no stable solution.

A potentially interesting area of research would be to look at the
possible tunnelling of these configurations.
From the worldvolume perspective, it is clear that tunnelling should
occur from the classically stable configuration at $r_H=r_-$ to some expanding solution.  On the
supergravity side, in the case with no dissolved charge,
instantons describing such a process exist, and have been described in
some detail by Dowker \emph{et al}~\cite{dowker:95}.  They correspond to the nucleation
of spherical branes within a fluxbrane.  It would be interesting to
look for analogous instantons in the case with dissolved charge.

\bigskip

\centerline{\bf Acknowledgements}

We would like to thank Miguel Costa, Bert Janssen and Simon Ross for
useful discussions and comments on a draft of this paper.  We are also indebted to Roberto Emparan for explaining
the stability arguments and urging us to look at the smeared
solutions.  DB is supported in part by the
EPSRC grant GR/N34840/01 and PMS by PPARC.


\begin{thebibliography}{10}

\bibitem{melvin:64}
M.~A. Melvin.
\newblock ``{P}ure {M}agnetic and {E}lectric {G}eons''.
\newblock {\em Phys. Lett.}, 8:65, 1964.

\bibitem{gibbons:88}
G.~W. Gibbons and K.~Maeda.
\newblock ``{B}lack {H}oles and {M}embranes in {H}igher {D}imensional
  {T}heories with {D}ilaton {F}ields''.
\newblock {\em Nucl. Phys.}, B448:741, 1988.

\bibitem{russo:94}
J.~G. Russo and A.~A. Tseytlin.
\newblock ``{C}onstant magnetic field in closed string theory: {A}n {E}xactly
  solvable model''.
\newblock {\em Nucl. Phys.}, B448:293, 1995.
\newblock hep-th/9411099.

\bibitem{russo:95}
J.~G. Russo and A.~A. Tseytlin.
\newblock ``{M}agnetic flux tube models in superstring theory''.
\newblock {\em Nucl. Phys.}, B461:131, 1996.
\newblock hep-th/9508068.

\bibitem{tseytlin:95}
A.~A. Tseytlin.
\newblock ``{C}losed superstrings in magnetic flux background''.
\newblock {\em Nucl. Phys. Proc. Suppl.}, 49:338, 1996.
\newblock hep-th/9510041.

\bibitem{russo:98}
J.~G. Russo and A.~A. Tseytlin.
\newblock ``{G}reen-{S}chwarz superstring action in a curved magnetic
  {R}amond-{R}amond background''.
\newblock {\em JHEP}, 9804:014, 1998.
\newblock hep-th/9804076.

\bibitem{emparan:96}
R.~Emparan.
\newblock ``{C}omposite black holes in external fields''.
\newblock {\em Nucl. Phys.}, B490:365, 1997.
\newblock hep-th/9610170.

\bibitem{chen:99}
C.~Chen, D.~V. Gal'tsov, and S.~A. Sharakin.
\newblock ``{I}ntersecting {M}-{F}luxbranes''.
\newblock {\em Grav. Cosmol.}, 5:45, 1999.
\newblock hep-th/9908132.

\bibitem{costa:00}
M.~S. Costa and M.~Gutperle.
\newblock ``{T}he {K}aluza-{K}lein {M}elvin {S}olution in {M}-theory''.
\newblock {\em JHEP}, 0103:027, 2001.
\newblock hep-th/0012072.

\bibitem{gibbons:01}
G.~W. Gibbons and C.~A.~R. Herdeiro.
\newblock ``{T}he {M}elvin {U}niverse in {B}orn-{I}nfeld {T}heory and other
  {T}heories of {N}on-{L}inear {E}lectrodynamics''.
\newblock {\em Class. Quant. Grav.}, 18:1677, 2001.
\newblock hep-th/0101229.

\bibitem{paul1}
P.~M. Saffin.
\newblock ``{G}ravitating {F}luxbranes''.
\newblock gr-qc/0104014.

\bibitem{gutperle:01}
M.~Gutperle and A.~Strominger.
\newblock ``{F}luxbranes in {S}tring {T}heory''.
\newblock hep-th/0104136.

\bibitem{russo:01}
J.~G. Russo and A.~A. Tseytlin.
\newblock ``{M}agnetic backgrounds and tachyonic instabilities in closed
  superstring theory and {M}-theory''.
\newblock hep-th/0104238.

\bibitem{sparks:01}
J.~F. Sparks.
\newblock ``{K}aluza-{K}lein {B}ranes''.
\newblock hep-th/0105209.

\bibitem{paul2}
P.~M. Saffin.
\newblock ``{F}luxbranes from p-branes''.
\newblock hep-th/0105220.

\bibitem{dowker:93}
F.~Dowker, J.~P. Gauntlett, D.~A. Kastor, and J.~Traschen.
\newblock ``{P}air {C}reation of {D}ilaton {B}lack {H}oles''.
\newblock {\em Phys. Rev.}, D49:2909, 1994.
\newblock hep-th/9309075.

\bibitem{dowker:95a}
F.~Dowker, J.~P. Gauntlett, G.~W. Gibbons, and G.~T. Horowitz.
\newblock ``{T}he {D}ecay of magnetic fields in {K}aluza-{K}lein theory''.
\newblock {\em Phys. Rev.}, D52:6929, 1995.
\newblock hep-th/9507143.

\bibitem{dowker:95}
F.~Dowker, J.~P. Gauntlett, G.~W. Gibbons, and G.~T. Horowitz.
\newblock ``{N}ucleation of {$P$}-{B}ranes and {F}undamental {S}trings''.
\newblock {\em Phys. Rev.}, D53:7115, 1996.
\newblock hep-th/9512154.

\bibitem{costa:01}
M.~S. Costa, C.~A.~R. Herdeiro, and L.~Cornalba.
\newblock ``{F}lux-branes and the {D}ielectric {E}ffect in {S}tring {T}heory''.
\newblock hep-th/0105023.

\bibitem{emparan:01}
R.~Emparan.
\newblock ``{T}ubular {B}ranes in {F}luxbranes''.
\newblock hep-th/0105062.

\bibitem{emparan:97}
R.~Emparan.
\newblock ``{B}orn-{I}nfeld {S}trings {T}unneling to {D}-branes''.
\newblock {\em Phys. Lett.}, B423:71, 1998.
\newblock hep-th/9711106.

\bibitem{myers:99}
R.~C. Myers.
\newblock ``{D}ielectric-{B}ranes''.
\newblock {\em JHEP}, 12:022, 1999.
\newblock hep-th/9910053.

\bibitem{taylor:99}
W.~Taylor and M.~Van Raamsdonk.
\newblock ``{M}ultiple {D}$p$-branes in {W}eak {B}ackground {F}ields''.
\newblock {\em Nucl. Phys.}, B573:703, 2000.
\newblock hep-th/9910052.

\bibitem{trivedi:00}
S.~P. Trivedi and S.~Vaidya.
\newblock ``{F}uzzy {C}osets and their {G}ravity {D}uals''.
\newblock {\em JHEP}, 0009:041, 2000.
\newblock hep-th/0007011.

\bibitem{douglas:95}
M.~R. Douglas.
\newblock ``{B}ranes within {B}ranes''.
\newblock hep-th/9512077.

\bibitem{callan:97}
C.~G. Callan and J.~M. Maldacena.
\newblock ``{B}rane {D}ynamics {F}rom the {B}orn-{I}nfeld {A}ction''.
\newblock {\em Nucl. Phys.}, B513:198, 1998.
\newblock hep-th/9708147.

\bibitem{gibbons:97}
G.~W. Gibbons.
\newblock ``{B}orn-{I}nfeld particles and {D}irichlet p-branes''.
\newblock {\em Nucl. Phys.}, B514:603, 1998.
\newblock hep-th/9709027.

\bibitem{leigh:89}
R.~G. Leigh.
\newblock ``{D}irac-{B}orn-{I}nfeld {A}ction from {D}irichlet {S}igma
  {M}odel''.
\newblock {\em Mod. Phys. Lett.}, A4:2767, 1989.

\bibitem{li:95}
M.~Li.
\newblock ``{B}oundary {S}tates of {D}-{B}ranes and {D}y-{S}trings''.
\newblock {\em Nucl. Phys.}, B460:351, 1996.
\newblock hep-th/9510161.

\bibitem{green:96}
M.~Green, J.~A. Harvey, and G.~Moore.
\newblock ``{I}-{B}rane {I}nflow and {A}nomalous {C}ouplings on {D}-{B}ranes''.
\newblock {\em Class. Qaunt. Grav.}, 14:47, 1997.
\newblock hep-th/9605033.

\bibitem{cvetic:96}
M.~Cveti\v{c} and D.~Youm.
\newblock ``{R}otating {I}ntersecting {M}-{B}ranes''.
\newblock {\em Nucl. Phys.}, B499:253, 1997.
\newblock hep-th/9612229.

\bibitem{csaki:99}
C.~Cs\'{a}ki, J.~Russo, K.~Sfetsos, and J.~Terning.
\newblock ``{S}upergravity {M}odels for $3+1$ {D}imensional {QCD}''.
\newblock {\em Phys. Rev.}, D60:044001, 1999.
\newblock hep-th/9902067.

\bibitem{hawking:95}
S.~W. Hawking and G.~T. Horowitz.
\newblock ``{T}he {G}ravitational {H}amiltonian, {A}ction, {E}ntropy, and
  {S}urface {T}erms''.
\newblock {\em Class. Quant. Grav.}, 13:1487, 1996.
\newblock gr-qc/9501014.

\bibitem{duff:96}
M.~J. Duff, H.~Lu, and C.~N. Pope.
\newblock ``{T}he {B}lack {B}ranes of {M}-theory''.
\newblock {\em Phys. Lett.}, B382:73, 1996.
\newblock hep-th/9604052.

\bibitem{vilenkin+shellard}
A.~Vilenkin and E.~P.~S. Shellard.
\newblock {\em ``{C}osmic {S}trings and other {T}opological {D}efects''}.
\newblock CUP, 1994.

\bibitem{costa:00b}
M.~S. Costa and M.~J. Perry.
\newblock ``{I}nteracting {B}lack {H}oles''.
\newblock {\em Nucl. Phys.}, B591:469, 2000.
\newblock hep-th/0008106.

\end{thebibliography}

\end{document}